\documentclass[12pt]{iopart} 
\usepackage{iopams}
\usepackage{graphicx}
\usepackage{color}
\usepackage{float}

\usepackage[normalem]{ulem} 

\newcommand{\ii}{\mathrm{i}} 

\newcommand{\imp}{ \mathrm{imp} }

\begin{document}

\title[Relative phase...]{Relative phase between $s_{\pm}$ superconducting order parameter components in a two-band model with impurities}

\author{V.A. Shestakov$^{1}$, M.M. Korshunov$^{1}$}
\address{$^1$Kirensky Institute of Physics, Federal Research Center KSC SB, Russian Academy of Sciences, Krasnoyarsk, Russia}
\ead{v\_shestakov@iph.krasn.ru}

\vspace{10pt}
\begin{indented}
\item[]\date{\today}
\end{indented}

\begin{abstract}
We obtain solutions for Eliashberg equations within the Nambu representation for a two-band model of iron-based superconductors with nonmagnetic impurities. Two cases of a transition between $s_{\pm}$ and $s_{++}$ states are considered: (i) the transition is accompanied by the abrupt change of the order parameter sign within one of the bands and (ii) the change is smooth. For both cases, we studied the role of a gauge defined by the coefficients preceding the Pauli matrices $\hat\tau_1$ and $\hat\tau_2$ in a self-energy expansion, which correspond to the components of the order parameter. We show that the absolute value of the order parameter is conserved for solutions in the clean and in the Born limits. In an intermediate case, between the Born and unitary limits, result depends on the solution for the clean limit. We show that a common gauge for the Eliashberg equations in which one of the order parameter components vanishes is essential for adequate description of the multiband superconducting systems. 
\end{abstract}

\vspace{2pc}



%

\section{Introduction}
The superconducting state, which is a fundamental quantum state in condensed media, has been first characterized qualitatively at the microscopic level within the Bardeen–Cooper–Schrieffer (BCS) theory \cite{bcs}.  However, theoretical breakthroughs in quantitative characterization of various aspects of this phenomenon are associated with the development of quantum field theory methods \cite{AGD} and the formulation of Eliashberg equations \cite{Eliashberg1960}. The Eliashberg theory provides an opportunity to analyse dynamic interaction, which is a step forward compared to the BCS theory where the interaction potential is static. The single-band approximation has long been considered completely applicable to many of the studied systems. Iron compounds (pnictides and chalcogenides) with a multisheeted Fermi surface are a typical counterexample. Certain solutions with different signs of the superconducting order parameter corresponding to different sheets are obtained in the spin-fluctuation mechanism of superconductivity sustained by Coulomb repulsion. If the  order parameter does not change its sign within a single sheet, the state is denoted as $s_{\pm}$ \cite{HirschfeldKorshunov2011}. If the order parameter has the same sign on all sheets of the Fermi surface, the state is $s_{++}$. An additional complication arises if impurity scattering is taken into account. Owing to the presence of several bands, intra- and interband types of this scattering may be distinguished \cite{Ohashi2002, Ohashi2004}. The description of states $s_{\pm}$ and $s_{++}$ requires at least two bands, which is the basis of minimal models \cite{KorshunovUFN2016}. At the same time, it was demonstrated that the presence of impurities may lead to a transition between states $s_{\pm}$ and $s_{++}$ in a two-band model \cite{EfremovKorshunov2011, KorshunovMagn2014}. Depending on the magnitude of generalized scattering cross section $\sigma$, this transition may be abrupt or proceed smoothly through a gapless state with the order parameter in one of the bands becoming zero before the sign change \cite{ShestakovKorshunovSUST2018}. This gapless state may manifest itself, e.~g., in variation of penetration depth of the magnetic field with an increase in the density of impurities or defects \cite{ShestakovKorshunovSUST2021}.

The standard  approach to the Eliashberg theory is as follows. In the Nambu representation \cite{Nambu}, the self-energy matrix specifying the Eliashberg equations is written as an expansion in Pauli matrices $\hat\tau_j$, where $j = 0, \ldots, 3$ and index ``0'' corresponds to unit matrix
\begin{equation}\label{eq1}
	\fl\hat{\Sigma}(\mathbf{k},\ii\omega_n) = \ii\omega_n[1-Z(\mathbf{k},\ii\omega_n)]\hat{\tau}_0 + \chi(\mathbf{k},\ii\omega_n)\hat{\tau}_3+\phi_1(\mathbf{k},\ii\omega_n)\hat{\tau}_1+\phi_2(\mathbf{k},\ii\omega_n)\hat{\tau}_2,
\end{equation}
where $\ii\omega_n = (2n + 1)i\pi T$ are the Matsubara frequencies; $n$ is an integer number; and $Z$, $\chi$, $\phi_1$, and $\phi_2$ are arbitrary real independent functions for which the system of Eliashberg equations is solved. Functions $Z$ and $\chi$ are associated with renormalization of Matsubara frequencies $\ii\omega_n$ and single-electron excitations of quasiparticles, respectively, due to the dynamic pairing interaction and effects related to the presence of impurities in the system. Functions $\phi_1$ and $\phi_2$ define the superconductor order parameter and superconducting gap
\begin{equation}\label{eq2}
	\Delta_{1,2}(\mathbf{k},\ii\omega_n) = \phi_{1,2}(\mathbf{k},\ii\omega_n) / Z(\mathbf{k},\ii\omega_n).
\end{equation}
If $\phi_1 = \phi_2 = 0$, a system is in its normal (non-superconducting) state. If $\phi_1 \neq 0$, $\phi_2 \neq 0$, or both functions assume non-zero values, the obtained solutions correspond to the superconducting state. Non-zero solutions for $\phi_1$ and $\phi_2$ have the following property: there are an infinite number of solutions for the pair of functions $\phi_1$ and $\phi_2$ differing by an arbitrary phase factor $\exp(\ii2\theta)$. Each of these pairs satisfies relation $\phi^2_1 + \phi^2_2 = const$ \cite{Nambu, allen}. In practice, a gauge is chosen such that the coefficient preceding matrix $\hat\tau_1$ (i.~e., $\phi_1$) or $\hat\tau_2$ (i.~e., $\phi_2$) is zero. This reduces the number of equations in the system and simplifies calculations. This gauge is valid for single-band superconductors (classical and unconventional) and multiband superconductors with electron–phonon interaction \cite{Nambu, allen}. However, it is far from obvious whether it remains valid in the case when the structure of the superconducting order parameter changes in a multiband system with a gapless state established in one of the bands.

In the present study, the solutions of the system of Eliashberg equations with both coefficients preceding Pauli matrices in self-energy expansion (\ref{eq1}) taken into account are examined within the two-band model of iron-based superconductors with nonmagnetic impurities. It is demonstrated that the absolute value of the order parameter is conserved, $\phi^2_{1\alpha n} + \phi^2_{2\alpha n} = const$, for solutions in the Born limit where the order  parameter changes sign abruptly in transition $s_{\pm} \to s_{++}$. In the case of an intermediate scattering strength with the order parameter sign change being smooth, the result depends on the gauge of the initial solution for the clean limit. The standard gauge for the Eliashberg equations with one of the functions ($\phi_{1an}$ or $\phi_{2an}$) being zero is not only acceptable as a means to reduce computational costs, but also necessary for stability and convergence of solutions.

\section{Model and method}
In the present study, we use the approach of $\xi$-integrated Green’s functions within the two-band model of iron-based superconductors \cite{KorshunovUFN2016}. A Green’s function is a matrix in band and Nambu spaces:
\begin{equation}\label{eq3}
\hat{\mathbf{g}}(\ii\omega_n) = \hat{g}_{\alpha}(\ii\omega_n) \otimes \mathbf{1}_{\alpha\beta},
\end{equation}
where
\begin{equation}\label{eq4}
	\hat{g}_{\alpha}(\ii\omega_n) = -\pi N_\alpha (\ii\tilde\omega_{\alpha n}\hat\tau_0 + \phi_{1\alpha n}\hat\tau_1 + \phi_{2\alpha n}\hat\tau_2) / Q_{\alpha n},
\end{equation}
$Q_{\alpha n} = \sqrt{\tilde\omega^2_{\alpha n} + \phi^2_{1\alpha n} + \phi^2_{2\alpha n}}$, $\ii\tilde\omega_{\alpha n} \equiv \ii\omega_nZ_{\alpha}(\ii\omega_n)$, $\phi_{1(2)\alpha n} \equiv \phi_{\alpha 1(2)}(\ii\omega_n)$, $N_\alpha$ is the density of quasiparticle states at the Fermi level in a normal metal in the band with index $\alpha = (a, b)$ and $\mathbf{1}_{\alpha\beta}$ is a unit matrix in the band space.

The self-energy in this approach is also a matrix of the same dimension as the Green’s function. It is also diagonal in band indices (in the general case, this is not true, but non-diagonal contributions may be neglected in the present analysis) and contains two contributions from superconducting interaction and from nonmagnetic impurity scattering 
\begin{equation}\label{eq5}
	\hat{\mathbf{\Sigma}}(\ii\omega_n) = \hat{\mathbf{\Sigma}}^{\mathrm{SC}}(\ii\omega_n) + \hat{\mathbf{\Sigma}}^{\mathrm{imp}}(\ii\omega_n).
\end{equation}
It is convenient to present self-energy expansion (\ref{eq1}) in Pauli matrices in the following form:
\begin{equation}\label{eq6}
	\hat{\Sigma}_{\alpha}^{\mathrm{SC}(\mathrm{imp})}(\ii\omega_n) = \sum_{j=0}^{2}\Sigma_{j\alpha}^{\mathrm{SC}(\mathrm{imp})}(\ii\omega_n)\hat\tau_j
\end{equation}

We assume that the contribution from superconducting interaction is produced primarily by the exchange of spin fluctuations and is repulsive in nature, but may also contain an additional attractive electron–phonon part. All these contributions are introduced into the self-energy via interaction functions $\lambda^\phi_{\alpha\beta}(n - n')$ and $\lambda^Z_{\alpha\beta}(n - n')$:
\begin{equation}\label{eq7}
	\Sigma_{0\alpha}^{\mathrm{SC}}(\ii\omega_n) = -\ii\pi T\sum_{\omega'_n,\beta}{ \lambda_{\alpha \beta}^{Z}(n-n') \tilde{\omega}_{\beta n'}/Q_{\beta n'}},
\end{equation}
\begin{equation}\label{eq8}
	\Sigma_{1(2)\alpha}^{\mathrm{SC}}(\omega_n) = -\pi T\sum_{\omega'_n,\beta}{ \lambda_{\alpha \beta}^{\phi}(n-n') \phi_{1(2)\beta n'}/Q_{\beta n'} },
\end{equation}
where
\begin{equation}\label{eq9}
	\lambda_{\alpha \beta}^{\phi,Z}(n-n') = 2\lambda_{\alpha \beta}^{\phi,Z}\int_{0}^{\infty}{ d\Omega \frac{ \Omega B(\Omega) }{ \left( \omega_n - \omega_{n'} \right)^2 + \Omega^2 } },
\end{equation}
is determined through coupling constants $\lambda^{\phi,Z}_{\alpha\beta}$ and normalized bosonic spectral function $B(\Omega)$, which characterizes the spectrum of spin excitations in the system \cite{KorshunovUFN2016}. The values of $\lambda^\phi_{\alpha\beta}$ specified by the contributions of Coulomb repulsion, spin fluctuations, and electron–phonon interaction may be either positive (attraction) or negative (repulsion), while the values of $\lambda^Z_{\alpha\beta}$ are always positive. For simplicity, $\lambda^Z_{\alpha\beta} = |\lambda^\phi_{\alpha\beta}| \equiv |\lambda_{\alpha\beta}|$ is often assumed.

The contribution of impurities is taken into account in the T-matrix approximation \cite{KorshunovUFN2016, ShestakovKorshunovSymmetry2018, ShestakovKorshunovSUST2018}, which yields the following expressions:
\begin{equation}\label{eq10}
	\fl\Sigma^{\mathrm{imp}}_{0a} = -\ii\Gamma_a\left[ \sigma(1-\eta^2)^2\tilde{\omega}_{an}/Q_{an}+(1-\sigma)\left(\eta^2N_a\tilde{\omega}_{an}/(N_bQ_{an}) + \tilde{\omega}_{bn}/Q_{bn}\right) \right] / (2\mathrm{D}_{\mathrm{imp}}),
\end{equation}
\begin{equation}\label{eq11}
	\fl\Sigma^{\mathrm{imp}}_{1(2)a} = \Gamma_a\left[ \sigma(1-\eta^2)^2\phi_{1(2)an}/Q_{an} + (1-\sigma)\left(\eta^2N_a\phi_{1(2)an}/(N_bQ_{an}) + \phi_{1(2)bn}/Q_{bn}\right) \right] / (2\mathrm{D}_{\mathrm{imp}}),
\end{equation}
where $\Gamma_a$ is the intensity of impurity scattering, which is proportional to impurity density $n_{\imp}$ and effective scattering cross section $\sigma$:
\begin{equation}\label{eq12}
	\Gamma_a = 2n_{\mathrm{imp}}\sigma /(\pi N_a),
\end{equation}
\begin{equation}\label{eq13}
	\sigma = \pi^2 u^2 N_a N_b / (1+\pi^2 u^2 N_a N_b),
\end{equation}
and the value of $\eta = v/u$ specifies the ratio between intraband ($v$) and inter-band ($u$) components of the scattering potential of impurities,
\begin{equation}\label{eq14}
	\mathrm{D}_{\mathrm{imp}} = (1-\sigma)^2 + \sigma^2(1-\eta^2)^2 + \sigma(1-\sigma)\kappa_{\mathrm{imp}},
\end{equation}
\begin{equation}\label{eq15}
	\fl\kappa_{\mathrm{imp}} = \eta^2(N_a^2 + N_b^2)/(N_aN_b) + 2(\tilde\omega_{an}\tilde\omega_{bn} + \phi_{1an}\phi_{1bn} + \phi_{2an}\phi_{2bn})/(Q_{an}Q_{bn}).
\end{equation}
Having inserted expressions (\ref{eq7}), (\ref{eq8}), (\ref{eq10}), and (\ref{eq11}) into formulae (\ref{eq6}) and (\ref{eq5}) and equated the resulting right-hand side of expression (\ref{eq5}) to the right-hand side of expansion (\ref{eq1}), we obtain the system of Eliashberg equations:
\begin{eqnarray}\label{eq16}
	&\tilde\omega_{an} = \omega_n + \pi T\sum_{\omega'_n,\beta}{ \lambda_{a \beta}^{Z}(n-n') \tilde{\omega}_{\beta n'}/Q_{\beta n'}} +\\
	&\fl+ \Gamma_a\left[ \sigma(1-\eta^2)^2\tilde{\omega}_{an}/Q_{an}+(1-\sigma)\left(\eta^2N_a\tilde{\omega}_{an}/(N_bQ_{an}) + \tilde{\omega}_{bn}/Q_{bn}\right) \right] / (2\mathrm{D}_{\mathrm{imp}}), \nonumber
\end{eqnarray}
\begin{eqnarray}\label{eq17}
	&\phi_{1a} = \pi T\sum_{\omega'_n,\beta}{\lambda_{a \beta}^{\phi}(n-n') \phi_{1\beta n'}/Q_{\beta n'}} +\\
	&\fl+ \Gamma_a\left[ \sigma(1-\eta^2)^2\phi_{1an}/Q_{an} + (1-\sigma)\left(\eta^2N_a\phi_{1an}/(N_bQ_{an}) + \phi_{1bn}/Q_{bn}\right) \right] / (2\mathrm{D}_{\mathrm{imp}}), \nonumber
\end{eqnarray}
\begin{eqnarray}\label{eq18}
	&\phi_{2a} = \pi T\sum_{\omega'_n,\beta}{\lambda_{a \beta}^{\phi}(n-n') \phi_{2\beta n'}/Q_{\beta n'}} +\\
	&\fl+ \Gamma_a\left[ \sigma(1-\eta^2)^2\phi_{2an}/Q_{an} + (1-\sigma)\left(\eta^2N_a\phi_{2an}/(N_bQ_{an}) + \phi_{2bn}/Q_{bn}\right) \right] / (2\mathrm{D}_{\mathrm{imp}}). \nonumber
\end{eqnarray}
One half of the set of Eliashberg equations for one band $a$ is presented here; the equations for band $b$ are obtained by paired substitution of band indices in these equations.

At first glance, Eqs. (\ref{eq17}) and (\ref{eq18}) are identical and should yield the same set of solutions. However, both equations include not only components ``1'' and ``2'' of the order parameter for one band, but also the order parameter for the second band via $Q_{\alpha n}$ denominators. Since the order parameter in one of the bands goes through zero and changes its sign in the transition between states $s_{\pm}$ and $s_{++}$ induced by nonmagnetic impurities, it is important to determine what kind of gauge is applicable to the family of solutions of such a system of Eliashberg equations.

\section{Results and discussion}
\begin{figure}[ht]
	\centering
	\includegraphics[width=0.75\textwidth]{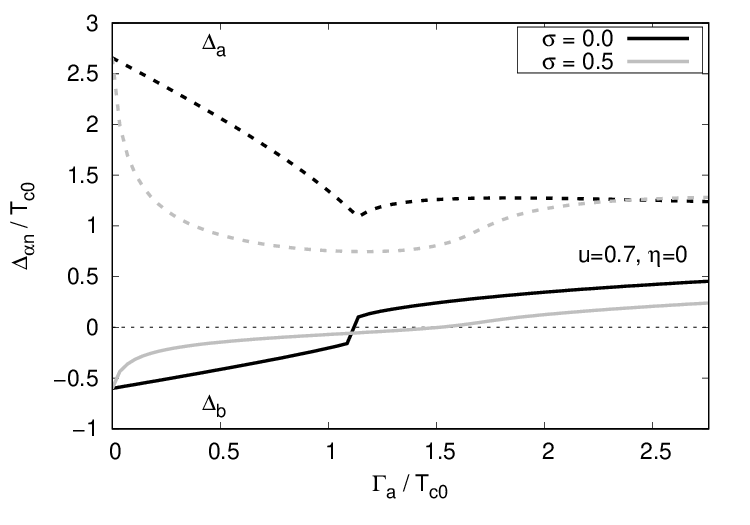}
	\caption{Plot of the dependence of the $\Delta_{\alpha n}$ gap function for the first Matsubara frequency ($n = 0$) on impurity scattering intensity $\Gamma_a$ for $\sigma = 0$ and $\sigma = 0.5$ at $\eta = 0$ and $T = 0.03T_{c0}$.}\label{fig1}
\end{figure}

In the clean limit, the values of coupling constants $(\lambda_{aa}, \lambda_{ab}, \lambda_{ba}, \lambda_{bb}) = (3.0, -0.2, -0.1, 0.5)$ used for calculations yield a superconducting state with the $s_{\pm}$ structure of the order parameter and a positive band-averaged coupling constant
\begin{equation*}
	\langle \lambda \rangle = [N_a(\lambda_{aa} + \lambda_{ab}) + N_b(\lambda_{ba}+\lambda_{bb})]/(N_a+N_b).
\end{equation*}
The critical temperature in the clean limit is $T_{c0} = 41.4$~K. Since it was demonstrated earlier \cite{ShestakovKorshunovSUST2018} that the presence of an intraband component in the impurity scattering potential does not affect qualitatively the phenomena under consideration, we assumed for simplicity that impurity scattering is non-existent in the intraband channel, $\eta = 0$. The $\sigma$ parameter may assume values ranging from zero in the Born limit for a weakly scattering impurity ($\pi u N_{a(b)} \ll 1$) to unity in the unitary limit of a strong scattering impurity potential ($\pi u N_{a(b)} \gg 1$). The unitary limit was not considered, since it follows from Eqs. (\ref{eq16})--(\ref{eq18}) \cite{KorshunovUFN2016} that the presence of nonmagnetic impurities does not affect the superconducting state in the unitary limit. Two cases differing in the nature of transition between the $s_{\pm}$ and $s_{++}$ states were chosen for examination: Born limit $\sigma = 0$ and intermediate case $\sigma = 0.5$. In the first case, the order parameter sign change in one of the bands (band $b$) is abrupt; in the second case, the sign change proceeds smoothly going through zero. Figure \ref{fig1} presents the variation of the order parameter in both bands with impurity scattering intensity $\Gamma_a$ for the cases of $\sigma = 0.0$ and $\sigma = 0.5$. The influence of strength of the impurity scattering potential, which is characterized by cross section $\sigma$, and relation $\eta$ between the intra- and interband components of the impurity potential were examined in more detail in \cite{ShestakovKorshunovSUST2018}. It was demonstrated that superconducting gaps change smoothly near the $s_{\pm} \to s_{++}$ transition at all values of $\sigma$ and $\eta$ except for the case of weak scattering with a small $\sigma$, where the smaller gap changes abruptly at the transition point and varies smoothly after that with an increase in $\Gamma_a$. The jump is evened out around $\sigma = 0.11$, and the transition becomes smooth. As the temperature grows, the abrupt behavior of the smaller gap changes to the smooth one at $T \approx 0.1 T_c$. This is why critical temperature $T_c$ has no singularities associated with the step-like nature of transition at small $\sigma$, remaining a smooth function of impurity scattering intensity $\Gamma_a$.

For further analysis, we present the pair of functions $\phi_{1\alpha n}$ and $\phi_{2\alpha n}$ as real and imaginary parts of complex function $\phi_{\alpha n} = \phi_{1\alpha n} + \ii\phi_{2\alpha n}$, which is written as
\begin{equation}\label{eq19}
	\phi_{\alpha n} = |\phi_{\alpha n}|\exp(\ii2\theta_\alpha) \equiv |\phi_{\alpha n}|[\cos(2\theta_\alpha) + \ii \sin(2\theta_\alpha)],
\end{equation}
where $|\phi_{\alpha n}| = \sqrt{\phi^2_{1\alpha n} + \phi^2_{2\alpha n}}$ is the modulus and $2\theta_\alpha$ is the phase of the superconducting order parameter in band $\alpha$.

At each step in $\Gamma_a$, the results from the previous step were taken as seed values of the renormalized Matsubara frequencies and the order parameter (starting from the clean limit) to solve the system of Eliashberg equations (\ref{eq16})--(\ref{eq18}) for a superconductor with impurities. This method allows one to obtain a solution for the superconducting phase without impurities with a fixed phase and modulus of the order parameter and trace the evolution of the system with successive introduction of impurities. The temperature was set to $T = 0.01T_{c0}$ in calculations.

\begin{figure}[p]
	\centering
	\includegraphics[width=0.7\textwidth]{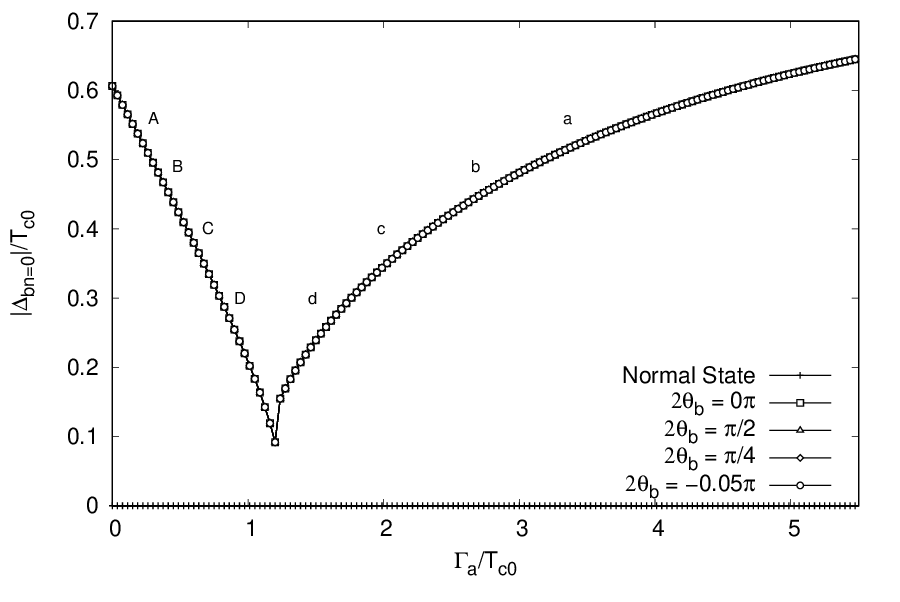}
	\caption{Dependences of the modulus of the $\Delta_bn$ gap function for the first Matsubara frequency ($n = 0$) on impurity scattering intensity $\Gamma_a$. The Born limit, $\sigma = 0$. The phase indicated for each plot corresponds to the phase of the solution in the clean limit ($\Gamma_a = 0$). Capital letters denote the branches of plots before the transition ($s_{\pm}$ state), while the branches after the transition ($s_{++}$ state) are denoted with lowercase letters. All plots (with the exclusion of the one for the normal state) match perfectly.}\label{fig2}
\end{figure}

\begin{figure}[p]
	\centering
	\includegraphics[width=0.7\textwidth]{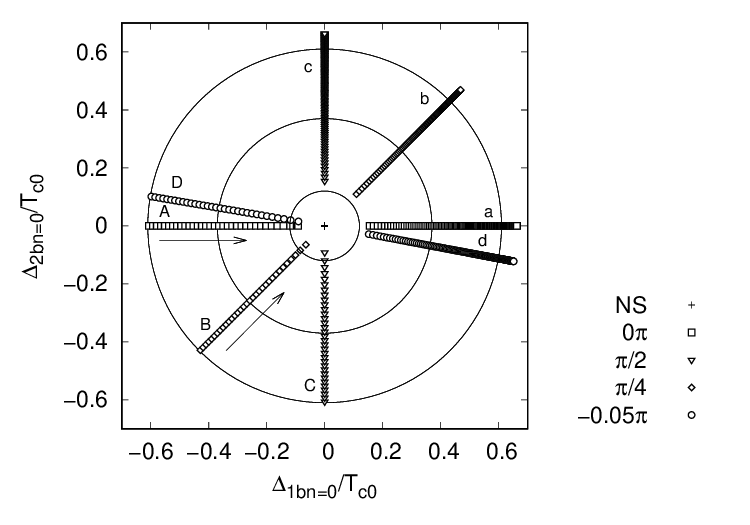}
	\caption{Values of the $\Delta_{bn}$ gap function for the first Matsubara frequency ($n = 0$) plotted on complex plane ($\Delta_{1b0}$, $\Delta_{2b0}$). The Born limit. Concentric circles indicate that the family of solutions for the superconducting gap for each fixed value lies on the $\Delta^2_{1b0} + \Delta^2_{2b0} = const$ circle. The outer circle corresponds to solutions for a clean superconductor. The phase indicated for each plot corresponds to the phase of the solution in the clean limit ($\Gamma_a = 0$). Capital letters denote the branches of plots before the transition ($s_{\pm}$ state), while the branches after the transition ($s_{++}$ state) are denoted with lowercase letters. Arrows indicate the direction in which $\Gamma_a$ increases.}\label{fig3}
\end{figure}

The calculation results for the Born limit are shown in Figures \ref{fig2} and \ref{fig3}. These figures present dependence $|\Delta_{b0}(\Gamma_a)|$ of the modulus of the complex superconducting gap function for the first Matsubara frequency ($n = 0$) on the impurity scattering intensity and the plot of function $\Delta_{b0}(\Gamma_a)$ itself on the complex plane in coordinates ($\Delta_{1b0}$, $\Delta_{2b0}$), respectively. The complex gap function is tied to the order parameter function by relation (\ref{eq2}). These figures demonstrate that the modulus of the gap function remains unchanged at fixed $\Gamma_a$ for all the presented families of solutions of the Eliashberg equations, and its phase changes by a fixed value of $\pi$ only at the moment of the $s_{\pm} \to s_{++}$ transition. Modulus $|\Delta_{b0}(\Gamma_a)|$ does not reach zero at the transition point (Figure \ref{fig2}), indicating that the gap changes abruptly.


\begin{figure}[p]
	\centering
	\includegraphics[width=0.7\textwidth]{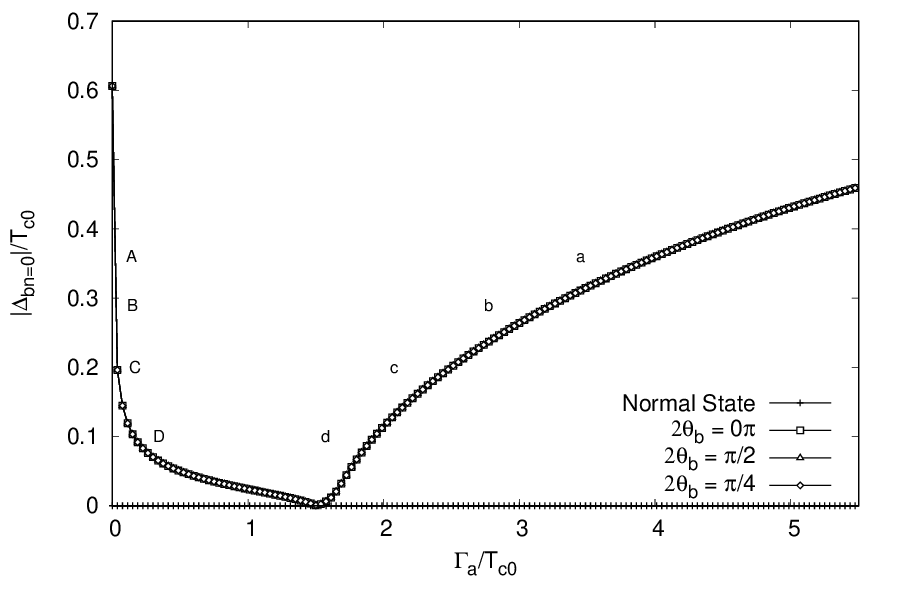}
	\caption{Dependences of the modulus of the $\delta_{bn}$ gap function for the first Matsubara frequency ($n = 0$) on impurity scattering intensity $\Gamma_a$. The intermediate case, $\sigma = 0.5$. In the 	presented solutions, the phase in the clean limit satisfies condition $2\theta_b = m\pi/4$. The phase indicated for each plot corresponds to the phase of the solution in the clean limit ($\Gamma_a = 0$).  Capital letters denote the branches of plots before the transition ($s_{\pm}$ state), while the branches after the transition ($s_{++}$ state) are denoted with lowercase letters. All plots (with the exclusion of the one for the normal state) match perfectly.}\label{fig4}
\end{figure}

\begin{figure}[p]
	\centering
	\includegraphics[width=0.7\textwidth]{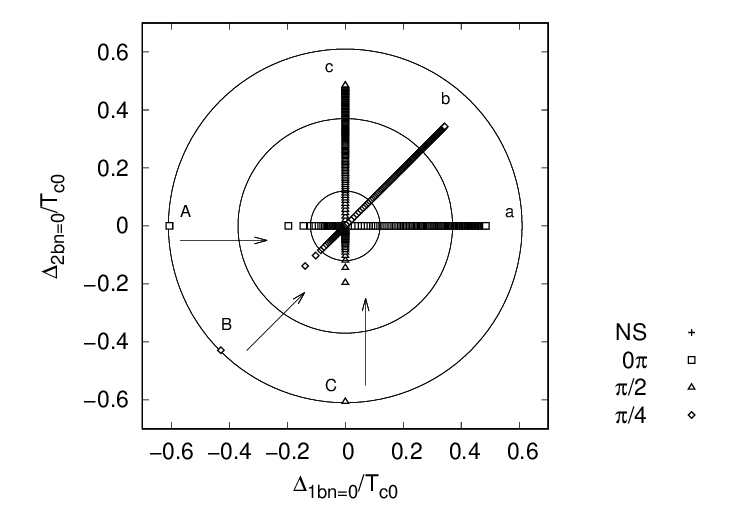}
	\caption{Values of the $\Delta_{b}n(\Gamma_a)$ gap function for the first Matsubara frequency ($n = 0$) plotted on complex plane ($\Delta_{1b0}$, $\Delta_{2b0}$) in directions $0$, $\pi/4$, and $\pi/2$. The intermediate case, $\sigma = 0.5$. As in Figure \ref{fig3}, the outer circle corresponds to solutions for a clean superconductor. Capital letters denote the branches of plots before the transition ($s_{\pm}$ state), while the branches after the transition ($s_{++}$ state) are denoted with lowercase letters. Arrows indicate the direction in which $\Gamma_a$ increases.}\label{fig5}
\end{figure}

\begin{figure}[p]
	\centering
	\includegraphics[width=0.7\textwidth]{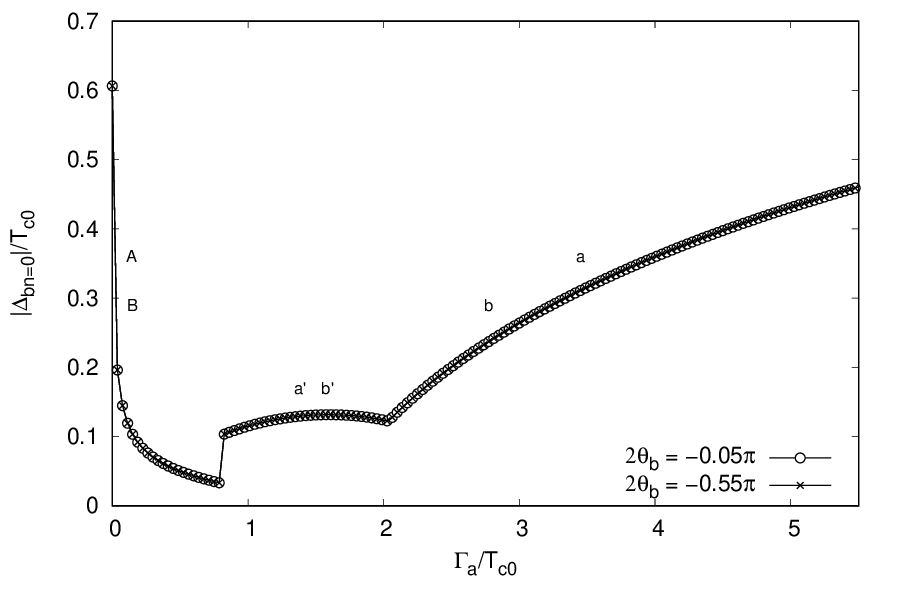}
	\caption{Dependences of the modulus of the 1bn gap function for the first Matsubara frequency ($n = 0$) on impurity scattering intensity $\Gamma_a$. The intermediate case, $\sigma = 0.5$. In the presented solutions, the phase in the clean limit does not satisfy condition $2\theta_b = m\pi/4$. The phase indicated for each plot corresponds to the phase of the solution in the clean limit ($\Gamma_a = 0$). Capital letters denote the branches of plots before the transition ($s_{\pm}$ state), while the branches after the transition ($s_{++}$ state) are denoted with lowercase letters.}\label{fig6}
\end{figure}

\begin{figure}[p]
	\centering
	\includegraphics[width=0.7\textwidth]{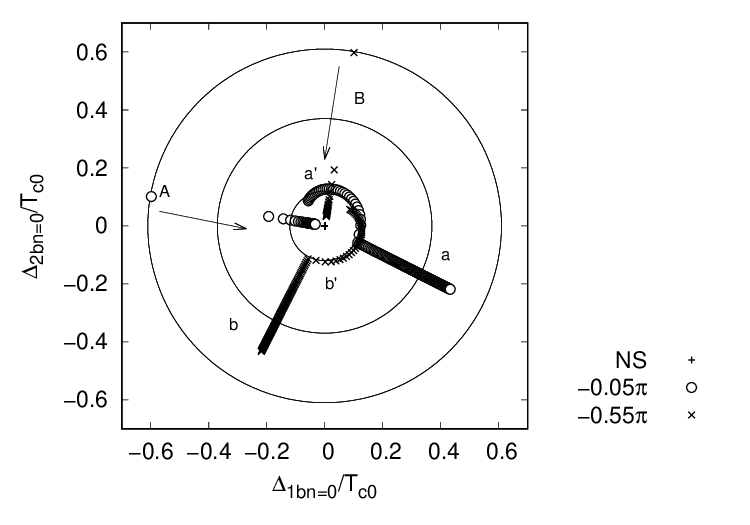}
	\caption{Values of the $\Delta_{bn}(\Gamma_a)$ gap function for the first Matsubara frequency ($n = 0$) plotted on complex plane ($\Delta_{1b0}$, $\Delta_{2b0}$) in directions differing from $\pi m/4$, where $m$ is an integer number. The intermediate case, $\sigma = 0.5$. As in Figure \ref{fig3}, the outer circle corresponds to solutions for a clean superconductor. Capital letters denote the branches of plots before the transition ($s_{\pm}$ state), while the branches after the transition ($s_{++}$ state) are denoted with lowercase letters. Primed lowercase letters indicate the regions of poor convergence of solutions of the Eliashberg equations. Arrows indicate the direction in which $\Gamma_a$ increases.}\label{fig7}
\end{figure}

A slightly different pattern is observed in the intermediate case with $\sigma = 0.5$ (Figures \ref{fig4}--\ref{fig7}). The solutions in which the phase in the clean limit is $2\theta_b = m\pi/4$, where $m$ is an integer number, behave as those in the Born limit. The sole difference is that the transition is smooth and the modulus of the gap function reaches zero (Figure \ref{fig4}); the phase changes by $\pi$ only at $|\Delta_{b0}| = 0$ (Figure \ref{fig5}). However, with just a small deviation from these symmetrical directions in plane ($\Delta_{1b0}$, $\Delta_{2b0}$), a region of poor convergence of the system of equations emerges near the $s_{\pm} \to s_{++}$ transition. This is seen in Figs. 6 and 7 in the cases with order parameter phases $2\theta_b = -0.05\pi$ and $2\theta_b = -0.55\pi$ for a clean superconductor in the region between $\Gamma_a = 0.8T_{c0}$ and $\Gamma_a = 2.1T_{c0}$, where the value of $|\Delta_{b0}|$ differs from the one corresponding to symmetric directions. The values of the $\Delta_{b0}$ modulus for these asymmetric directions are plotted in Figure \ref{fig6} for clarity, since the procedure of numerical solution of the system of equations entered an infinite loop and stopped at the limit of iterations, leaving the condition on the residual of the solution unfulfilled. However, it is worth noting that the modulus of the gap function for a fixed value of $\Gamma_a$ remains unchanged in each case of this kind. As for the phase of such a solution, it can be seen from Figure \ref{fig7} that it is unstable in this region of poor convergence: the solution ``wanders'' along complex plane ($\Delta_{1b0}$, $\Delta_{2b0}$) until a new stable family of solutions is reached. The difference between the order parameter phases for different bands remains fixed: $|2\theta_a - 2\theta_b| = \pi$ in state
$s_{\pm}$ and $|2\theta_a - 2\theta_b| = 0$ in state $s_{++}$.


This behavior of solutions at $\sigma = 0.5$ outside of symmetric directions is attributable to the fact that the phase of the gap function (order parameter) in band $b$ is undefined at near-zero $|\Delta_{b0}|$ values, and the system of Eliashberg equations cannot converge to a stable solution. In the Born limit, any direction on the ($\Delta_{1b0}$, $\Delta_{2b0}$) plane yields a stable solution, since the $s_{\pm} \to s_{++}$ transition is abrupt for near-zero values of $\sigma$ and the $\Delta_{b0}$ function does not have enough time to reach such values at which its phase is undefined. In the case of symmetric directions $2\theta_b = 0 \pm \pi m$ and $2\theta_b = \pi/2 \pm \pi m$, the gap function is real (the equations are solved either for the real or for the imaginary part of $\Delta_{bn}$ only), and its phase is actually defined at any value of $\Gamma_a$. A similar pattern is seen in the $2\theta_b = \pi/4 \pm \pi m$ direction, where Eqs. (\ref{eq17}) and (\ref{eq18}) match (except for sign), and their solutions also turn out to be stable.

\section{Conclusion}
The solutions of Eliashberg equations in the Nambu representation for a two-band model of iron-based superconductors with nonmagnetic impurities were studied numerically. The behavior of solutions in the presence of a transition between the $s_{\pm}$ and $s_{++}$ states with the coefficients preceding Pauli matrices $\hat\tau_1$ and $\hat\tau_2$ in the self-energy expansion being non-zero in the original system was examined. The modulus of the order parameter is preserved in the solutions for a clean superconductor; i.~e., the family of solutions satisfies condition $\phi^2_{1\alpha n} + \phi^2_{2\alpha n} = const$. This is also true in the case of introduction of nonmagnetic impurities into the system in Born limit $\sigma = 0$, where the order parameter changes sign abruptly during the transition. Far from the Born limit ($\sigma = 0.5$, where the order parameter changes smoothly), when impurities are introduced systematically into the system, the result depends on the solution obtained in the clean limit. Specifically, if the solutions of Eliashberg equations for a superconductor without impurities satisfy gauges $\phi_{1\alpha n} \neq 0$ and $\phi_{2\alpha n} = 0$, $\phi_{1\alpha n} = 0$ and $\phi_{2\alpha n} \neq 0$ or $\phi_{1\alpha n} = \phi_{2\alpha n} \neq 0$, the modulus of the order parameter also remains unchanged at each step in $\Gamma_a$ . In all the other families of solutions in the clean limit, a region of poor convergence of solutions of the equations emerges when impurities are introduced into the system, and it is impossible to tell for certain how the order parameter behaves in such cases. It follows  from the above that the standard gauge for the Eliashberg equations, where one of the functions $\phi_{1\alpha n}$ or $\phi_{2\alpha n}$ is identically equal to zero, is not only acceptable as a means to reduce computational costs, but also necessary for stability and convergence of solutions for a superconductor undergoing a transition with a change of sign of the order parameter.

\ack
This study was carried within the state assignment of Kirensky Institute of Physics.
\newline

\noindent{\textbf{Conflict of interest}}\newline

\noindent{The authors declare that they have no conflict of interest.}

\section*{References}

\bibliographystyle{iopart-num}
\bibliography{FTT2024_VAS_1col_arxiv}

\end{document}